\documentclass[twocolumn,showpacs,aps,prd,amsmath,amssymb,amsfonts%
]{revtex4-1}
\usepackage{graphicx}
\usepackage{bm}
\usepackage{relsize}

\begin{document}
\newcommand*{\bcp}{\ensuremath{B_c^+}}
\newcommand*{\vcd}{\ensuremath{V_{cd}}}
\newcommand*{\vcs}{\ensuremath{V_{cs}}}
\newcommand*{\bsn}{\ensuremath{B_s^0}}
\newcommand*{\bn}{\ensuremath{B^0}}
\newcommand*{\dstp}{D^{*+}}
\newcommand*{\dsstp}{D_s^{*+}}
\newcommand*{\beast}{\begin{eqnarray*}}
\newcommand*{\eeast}{\end{eqnarray*}}  
\newcommand*{\bea}{\begin{eqnarray}}   
\newcommand*{\eea}{\end{eqnarray}}     
\newcommand*{\bfr}{\mathcal{B}}        
\newcommand*{\br}[2]{\frac{\mathcal{B}(\bcp\ra #1)}{\mathcal{B}(\bcp\ra #2)}}
\newcommand*{\qratio}[4]{\sqrt{\frac{\lambda(m_{\bcp}^2,m_{#1}^2,m_{#2}^2)}
{\lambda(m_{\bcp}^2,m_{#3}^2,m_{#4}^2)}} }
\newcommand*{\rsq}[2]{\left|\frac{#1}{#2}\right|^2}
\newcommand*{\rf}[1]{(\ref{#1})}
\newcommand*{\bmln}{\begin{multline}}
\newcommand*{\emln}{\end{multline}}  
\newcommand*{\eg}{\textit{e.g.}}     
\newcommand*{\nl}{\nonumber \\}
\newcommand*{\bi}{\bibitem}
\newcommand*{\be}{\begin{equation}}
\newcommand*{\ee}{\end{equation}}
\newcommand*{\ra}{\rightarrow}
\newcommand*{\dek}[1]{\times10^{#1}}
\newcommand*{\bt}{\begin{table}}       
\newcommand*{\et}{\end{table}}         
\newcommand*{\btab}{\begin{tabular}}   
\newcommand*{\etab}{\end{tabular}}    
\newcommand*{\ea}{\textit{et al.}}
\newcommand*{\phrd}[3]{Phys.~Rev.~D~\textbf{#1}, #2 (#3)}
\newcommand*{\phrl}[3]{Phys.~Rev.~Lett.~\textbf{#1}, #2 (#3)}
\newcommand*{\pr}[3]{Phys.~Rev.~\textbf{#1}, #2 (#3)}      
\newcommand*{\npbps}[3]{Nucl.~Phys.~B (Proc. Suppl.) \textbf{#1}, #2 (#3)}  
\newcommand*{\rmph}[3]{Rev.~Mod.~Phys.~\textbf{#1}, #2 (#3)}
\newcommand*{\ibid}[3]{\textit{ ibid.} \textbf{#1}, #2 (#3)}
\newcommand*{\epjc}[3]{Eur. Phys. J. C \textbf{#1}, #2 (#3)}
\newcommand*{\jhep}[3]{JHEP \textbf{#1}, #2 (#3)}
\newcommand*{\phrp}[3]{Phys. Rep. \textbf{#1}, #2 (#3)}
\def\babar{\mbox{\slshape B\kern-0.1em{\smaller A}\kern-0.1em
    B\kern-0.1em{\smaller A\kern-0.2em R}}}

\title{Meson dominance implications for the bottomness preserving decays 
of the \bm{$B_c^+$} meson}

\author{Peter Lichard}
\affiliation{
Institute of Physics, Silesian University in Opava, 746 01 Opava, 
Czech Republic\\
and\\
Institute of Experimental and Applied Physics, Czech Technical University 
in Prague, 128 00 Prague, Czech Republic
}

\begin{abstract}
Encouraged by a recent observation of the $B_c^+\ra B_s^0 + \pi^+$ decay 
by the LHCb collaboration we present the meson dominance predictions for 
other weak decays of the $B_c^+$ into $B_s^0$ or $B^0$ in the form of 
branching ratios to the observed decay.
\end{abstract}


\maketitle

The LHCb collaboration at the CERN LHC collider has recently announced
\cite{lhcb} the observation of the bottomness--preserving (BP) decay 
$B_c^+\ra B_s^0 + \pi^+$ with significance in excess of five standard 
deviations independently in two decay channels of \bsn. This sets a hope 
that similar BP decays of the \bcp~meson may be found either in the 
already accumulated data or after the LHC reopens with higher energy 
and luminosity in 2015. 

In this note we present the estimates of the branching ratios of other
BP decays of the \bcp~relative to the already observed one. We use 
the meson dominance (MD) model \cite{md}, which describes well \cite{pl} 
the meson decays that fall into the ``external W-emission'' category 
according to the quark-diagram nomenclature \cite{chau83}.

\begin{figure}[b]
\includegraphics[width=6cm,height=1.8cm]{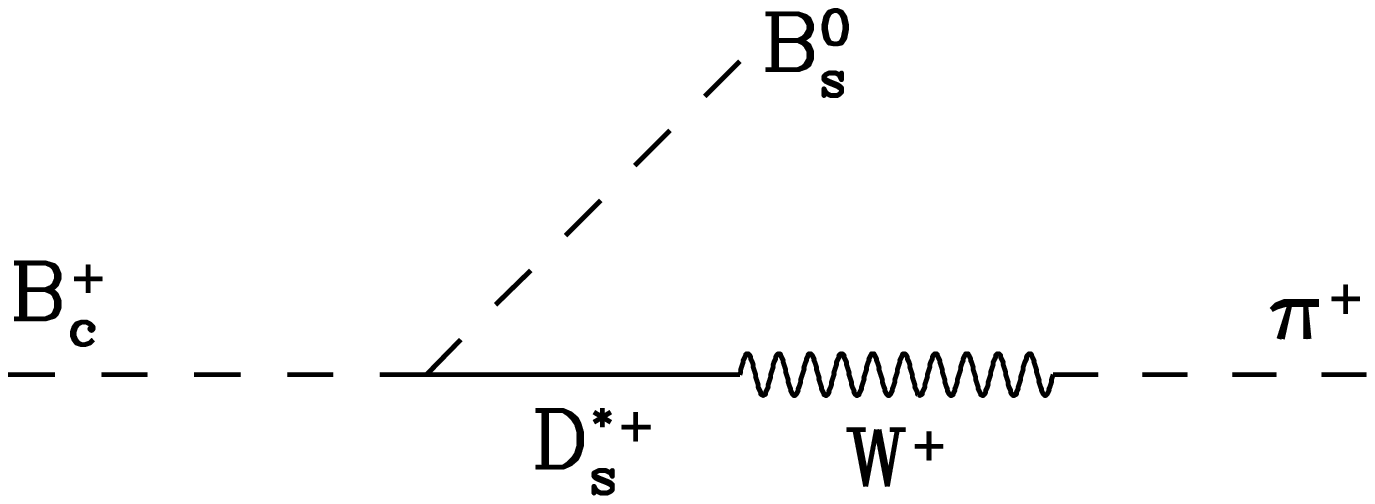}   
\caption{\label{fig:bcpbs0pip}Decay $\bcp\ra\bsn+\pi^+$ in the MD model.}
\end{figure}

The decay of the \bcp~(or, to be general, $P_1$) into the neutral meson
\bsn~($P_2$) and the $\pi^+$ ($P_3$) is described by a diagram 
depicted in Fig. \ref{fig:bcpbs0pip}. The diagram contains first a strong 
interaction vertex entered by $P_1$, $P_2$, and a positive vector meson $V$ 
(here $\dsstp$) with flavor quantum numbers required by the conservation laws. 
The vector meson then couples to the gauge boson $W^+$, which in turn couples 
to the outgoing pseudoscalar meson $P_3$ (here $\pi^+$). The corresponding 
partial decay width is given by 
\be
\label{pspspsga}
\Gamma(P_1\ra P_2+P_3)=\frac{G_F^2X_{P_1P_2V}Z_{P_3}}
{16\pi m_1^3} 
(x-y)^2 \lambda^{1/2}(x,y,m_3^2)\ ,
\ee
where $x=m_1^2$, $y=m_2^2$, and function $\lambda$ is defined by
$\lambda(x,y,z)=x^2+y^2+z^2-2xy-2xz-2yz$. The definition of the
dimensionless parameter $X_{P_1P_2V}$
\[
X_{P_1P_2V}=\left|w_VV_V\frac{g_{VP_1P_2}}{g_\rho}\right|^2
\]
includes the strong coupling constant $g_{VP_1P_2}$, the $\rho\pi\pi$
coupling constant $g_\rho$, the element $V_V$ of the CKM matrix pertinent 
to valence quark and antiquark of the meson $V$, and a dimensionless 
parameter $w_V\approx 1$ characterizing the deviation of the $V-W^+$ 
coupling from the $\rho^+-W^+$ one. At present, there is no chance of
getting the value of $X_{\bcp\bsn\dsstp}$ from its components. But as
we are going to relate the branching fractions of the various $\bcp\ra\bsn$
transitions, this unknown quantity will cancel.

The parameter $Z_{P}$, which is another important ingredient of
Eq.~\rf{pspspsga}, is defined in terms of the pseudoscalar decay constant
$f_P$ and the CKM matrix element $V_P$ corresponding to the valence quark
composition of a particular pseudoscalar meson $P$ by
\be
\label{zp}
Z_P=\left|f_P V_P\right|^2.
\ee
Its value is determined from the muonic decay width given by the formula
\cite{md}
\begin{multline*}
\Gamma(P\ra\mu^++\nu_\mu)=\frac{G_F^2}{8\pi}Z_Pm_\mu^2m_P
\left(1-\frac{m_\mu^2}{m_P^2}\right)^2\\
\times\left[1+\mathcal{O}(\alpha)\right].
\end{multline*}
We have included the radiative corrections
$\mathcal{O}(\alpha)$ following \cite{radcorr2}. The results obtained
by using the experimental muonic decay widths
\cite{pdg2012} are shown in Table~\ref{tab:zp}. 
\bt[b]
\caption{Parameters $Z_{P}$ characterizing the coupling
of pseudoscalar mesons to the charged gauge boson and their sources.
For definition, see Eq.~\rf{zp}.}
\label{tab:zp}
\btab{lcc}
P & $Z_{P}$ (MeV$^2$) & Source \\
\hline
$\pi^+$ & $(1.6158\pm0.0019)\!\times\!10^4$ & $\pi^+\ra \mu^+\nu_\mu$\\
$K^+$ & $(1.2307\pm0.0033)\!\times\!10^3$   & $K^+\!\ra \mu^+\nu_\mu$
\etab
\et

Now we are ready to calculate the branching ratio
\begin{multline*}
\br{\bsn+K^+}{\bsn+\pi^+}=\frac{Z_{K^+}}{Z_{\pi^+}}\qratio{\bsn}{K^+}
{\bsn}{\pi^+}\\
=(6.470\pm0.025)\dek{-2},
\end{multline*}
which shows that the decay $\bcp\ra\bsn+K^+$ will not be probably observed
soon.

\begin{figure}[h]
\includegraphics[width=6cm,height=2.5cm]{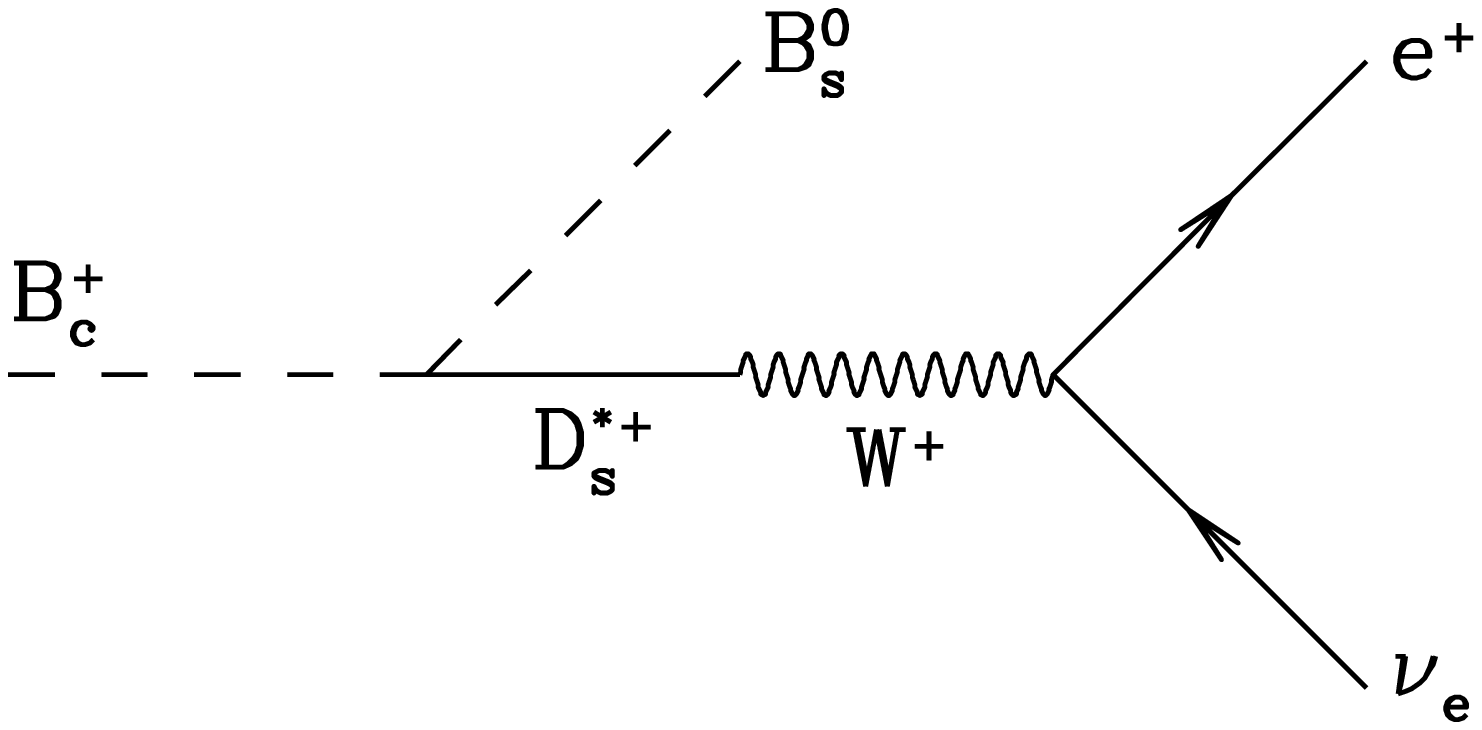}  
\caption{\label{fig:pspslnu}The semileptonic decay of the \bcp~to 
\bsn~meson in the MD model.}
\end{figure}

Let us turn now to the semileptonic decays. The decay $\bcp~(P_1)\ra 
\bsn~(P_2)+\ell^++\nu_\ell$ is in the MD model described by
diagram in Fig.~\ref{fig:pspslnu}. The differential decay width in $t$,
which is the square of the four-momentum transfer from $P_1$ to $P_2$, 
is given by
\begin{multline}
\label{dpspslnu}
\frac{d\Gamma}{dt}=\frac{G_F^2X_{P_1P_2V}}
{3(4\pi m_1)^3}\ \frac{(t-z)^2}{t^3}\ \lambda^{1/2}
(x,y,t)\left(\frac{r}{r-t}\right)^2\\
\times\!\left[(2t+z)\lambda(x,y,t)+\frac{3z}{r^2}(x-y)^2
(r-t)^2
\right].
\end{multline}
Newly defined parameters are $r=m_V^2$ and $z=m_\ell^2$. Formula
\rf{dpspslnu} is a simplified but equivalent version of Eq.~(4.3)
from Ref. \cite{md}. 
After factorizing out the unknown multiplication constant we can
perform the numerical integration of \rf{dpspslnu} and get the total 
semileptonic decay width (up to that multiplication constant). After 
dividing it by \rf{pspspsga}, the unknown $X$-factor cancels and we
are getting
\[
\br{\bsn+e^++\nu_e}{\bsn+\pi^+}=0.392\pm0.006
\]
and
\[
\br{\bsn+\mu^++\nu_e}{\bsn+\pi^+}=0.367\pm0.006.
\]

We can also compare the branching fractions of the modes with different
$B$-mesons in the final states, \eg,
\begin{multline*}
\br{\bn+\pi^+}{\bsn+\pi^+}=R_X\frac{m^2_{\bcp}-m^2_{\bn}}
{m^2_{\bcp}-m^2_{\bsn}} \\
\times\qratio{\bn}{\pi^+}{\bsn}{\pi^+},
\end{multline*}
where
\[
R_X=\frac{X_{\bcp\bn\dstp}}{X_{\bcp\bsn\dsstp}}
\]
is an unknown quantity. We can get a crude estimate of its value if assuming
that the light flavor SU(3) symmetry is not badly broken. Then we can write
\[
R_X\approx \rsq{\vcd}{\vcs}\approx 5.3\dek{-2}
\] 
and
\[
\br{\bn+\pi^+}{\bsn+\pi^+}\approx 7\dek{-2}.
\]
The semileptonic transitions of \bcp~to \bn~will be suppressed relative to
those to \bsn~by the same factor.

To conclude: The semileptonic decays of the \bcp~to \bsn~meson are, based
on the magnitude of their branching fractions, the best candidates for the
experimental observation. The branching fraction of $\bcp\ra 
\bsn+ e^++\nu_e$ ($\bcp\ra \bsn+\mu^++\nu_\mu$) is about thirty-nine
(thirty-seven) per cent of that of the already observed decay $B_c^+\ra B_s^0
+ \pi^+$. 

I am indebted to Dr. Tim Gershon for useful correspondence and to Dr. 
Josef Jur\'{a}\v{n} for discussions. This work was supported by the 
Research Program MSM6840770029 of the Czech Ministry of Education, 
Youth and Sports.

\end{document}